\documentclass[amssymb,nobibnotes,nofootinbib,aps,preprintnumbers,prd]{revtex4}

 \usepackage{here}

\usepackage[dvipdfmx]{graphicx}
\usepackage{amssymb,amsfonts,amsmath,bm,color,multirow,cases,empheq,hyperref}
\usepackage{mathrsfs}
\definecolor{darkergreen}{rgb}{0,0.5,0}

\usepackage{enumerate}
\usepackage{ulem}
\usepackage{afterpage}

\hypersetup{%
 setpagesize=false,
 bookmarksnumbered=true,%
 bookmarksopen=true,%
 colorlinks=true,%
 linkcolor=blue,%
 citecolor=red}

\newcommand{\fr}[1]{\frac{1}{#1}}

\newcommand{\nonum}{\nonumber\\ }

\newcommand{\cout}[1]{}

\newcommand{\arrayL}[1]{\left(\begin{array}{#1}}
\newcommand{\arrayR}{\end{array}\right)}
\newcommand{\arrayLb}[1]{\left[\begin{array}{#1}}
\newcommand{\arrayRb}{\end{array}\right]}

\begin{document}
\title{
Exploring non-supersymmetric black holes with multiple bubbles\\ in five-dimensional minimal supergravity
}

\author{Ryotaku Suzuki}
\email{sryotaku@toyota-ti.ac.jp}
\author{Shinya Tomizawa}
\email{tomizawa@toyota-ti.ac.jp}
\affiliation{\vspace{3mm}Mathematical Physics Laboratory, Toyota Technological Institute\vspace{2mm}\\Hisakata 2-12-1, Tempaku-ku, Nagoya, Japan 468-8511\vspace{3mm}}

\begin{abstract}

The topological censorship theorem suggests that higher dimensional black holes can possess the domain of outer communication (DOC) of nontrivial topology.
In this paper, we seek for a  black hole 
coexisting with two bubbles  adjacent to the horizon in five-dimensional minimal supergravity, under the assumptions of stationarity and bi-axisymmetry. For simplicity, 
we also assume that the spacetime is symmetric under the exchange of the two axisymmetric Killing vectors.
To find the solution, we combine the inverse scattering method and the Harrison transformation,
and  we  present the conditions for the absence of conical, orbifold and Dirac-Misner string singularities, respectively.
As the result, we find that the black hole with topology of $S^3$ or $S^2\times S^1$ can be supported by two bubbles if we admit the conical singularities (deficits).

\end{abstract}

\date{\today}
\preprint{TTI-MATHPHYS-33}

\maketitle

\section{Introduction}

In four dimensions, the topological censorship theorem 
constrains the topology of the domain of outer communication (DOC) to be the trivial one such that ${\rm DOC}\cap \Sigma={\mathbb R}^3\setminus {\mathbb B}^3 \ (\Sigma: {\rm timeslice})$~\cite{Friedman:1993ty}. 
Based on this theorem, the uniqueness theorem for black holes is established, stating that the Kerr-Newman solution~\cite{Newman:1965my} uniquely describes the only asymptotically flat, stationary black hole in Einstein gravity, at most, including the Maxwell field~\cite{Mazur:1982db,Bunting1983}.
However, in higher dimensions, the topological censorship theorem admits the spacetime having homology groups of higher rank, leading to a certain nontrivial DOC topology. 
For example, in five-dimensions, with the further assumption of biaxisymmetry, it has been shown that the DOC can have the nontrivial topologies of $[{\mathbb R}^4 \# m (S^2 \times S^2 )\# 
n {\mathbb C}{\mathbb P}^2\# n'(-{\mathbb C}{\mathbb P}^2)]\setminus {\mathbb B}^4$ $(m,n,n' =0,1,2,\ldots)$~\cite{Hollands:2010qy}.
This suggests infinite non-uniqueness even among spherical black holes, and one might expect the existence of many types of spherical black holes with nontrivial DOC topology, beyond the known spherical black holes (such as the Myers-Perry solution~\cite{Myers:1986un} in  five-dimensional Einstein theory and  Cvetic-Youm solutions~\cite{Cvetic:1996xz} in five-dimensional minimal supergravity) with a trivial DOC topology (${\rm DOC}\cap \Sigma = {\mathbb R}^4 \setminus {\mathbb B}^4$). In fact, since 
uniqueness theorems~\cite{Morisawa:2004tc,Tomizawa:2009ua} for these examples assume the trivial topology of DOC, it is plausible that such solutions may exist.

\medskip

In the five-dimensional vacuum Einstein theory, however, 
such a solution has not been found yet despite past attempts (for example, Ref.~\cite{tomizawa:2015talk}).
The first example of such black hole solutions have been found by Kunduri and Lucietti as a BPS (Bogomol'nyi-Prasad-Sommerfield) solution within the bosonic sector of five-dimensional minimal supergravity, which has the DOC topology of $[{\mathbb R}^4 \# (S^2 \times S^2) ]\setminus {\mathbb B}^4$~\cite{Kunduri:2014iga}.
This is generalized to more generic BPS black holes in Ref.~\cite{Breunholder:2017ubu}.
On the other hand,  in our previous works~\cite{Suzuki:2023nqf,Suzuki:2024phv,Suzuki:2024abu}, the non-BPS black holes having the DOC topology of $[{\mathbb R}^4 \# {\mathbb C}{\mathbb P}^2 ]\setminus {\mathbb B}^4$ have been constructed by combining the inverse scattering method (ISM)~\cite{Belinsky:1979mh,Belinski:2001ph,Pomeransky:2005sj} and duality transformations of five-dimensional minimal supergravity (the Harrison~\cite{Bouchareb:2007ax} and Ehlers~\cite{Giusto:2007fx} transformations).
Because these non-BPS black holes have the BPS limit only to the spherical black hole with the trivial DOC topology (BMPV black hole~\cite{Breckenridge:1996is}), there are missing links between the known BPS  and non-BPS solutions.
Moreover, these black holes are decorated by $2$-cycles, often called as {\it bubbles}, outside the horizon.
Up to now, while the solution with more than two bubbles are found for BPS black holes, only a single bubble solution is found for non-BPS black holes.

\medskip

In this paper, we aim to investigate the existence of the non-BPS black hole solution having two bubbles which includes the non-BPS extension of  the result in Ref.~\cite{Kunduri:2014iga}.
 We study the solution for an asymptotically flat, stationary, bi-axisymmetric,  black hole in five-dimensional minimal supergravity, whose action is given by
 \begin{eqnarray}
S=\frac{1}{16 \pi G_5}  \left[ 
        \int d^5x \sqrt{-g}\left(R-\frac{1}{4}F^2\right) 
       -\frac{1}{3\sqrt{3}} \int F\wedge F\wedge A 
  \right] \,, 
\label{action} 
\end{eqnarray} 
where $F=dA$. 
The field equations of this action is the Einstein equation and the Maxwell equation with a Chern-Simons term, which are expressed as 
\begin{eqnarray}
 R_{\mu \nu } -\frac{1}{2} R g_{\mu \nu } 
 = \frac{1}{2} \left( F_{\mu \lambda } F_\nu^{ ~ \lambda } 
  - \frac{1}{4} g_{\mu \nu } F_{\rho \sigma } F^{\rho \sigma } \right) \,, 
 \label{Eineq}
\end{eqnarray}
and 
\begin{eqnarray}
 d\star F+\frac{1}{\sqrt{3}}F\wedge F=0 \,. 
\label{Maxeq}
\end{eqnarray}
Under the existence of one timelike Killing vector $\xi_0 = \partial/\partial t$ and one spacelike axial Killing vector $\xi_1 = \partial/\partial \psi$, this theory reduces to the $G_{2(2)}/[SL(2,{\mathbb{R}}) \times SL(2,{\mathbb{R}})]$ non-linear sigma models coupled with three-dimensional gravity~\cite{Mizoguchi:1998wv,Mizoguchi:1999fu}. 
Also assuming a third spacelike axial Killing vector $\xi_2 = \partial/\partial \phi$ commuting with the other two, the theory becomes the $G_{2(2)}/[SL(2,\mathbb{R})\times SL(2,\mathbb{R})]$ nonlinear sigma model.
Utilizing the invariance of the theory under the action of the $G_{2(2)}$-group, one can produce a new solution from a known solution by applying a $G_{2(2)}$-transformation.
In this paper, we use the Harrison transformation~(Eq.~(119) in Ref.~\cite{Bouchareb:2007ax}) to produce the charged solution in five-dimensional supergravity from the five-dimensional asymptotically flat vacuum solution.

\begin{figure}
\includegraphics[width=8cm]{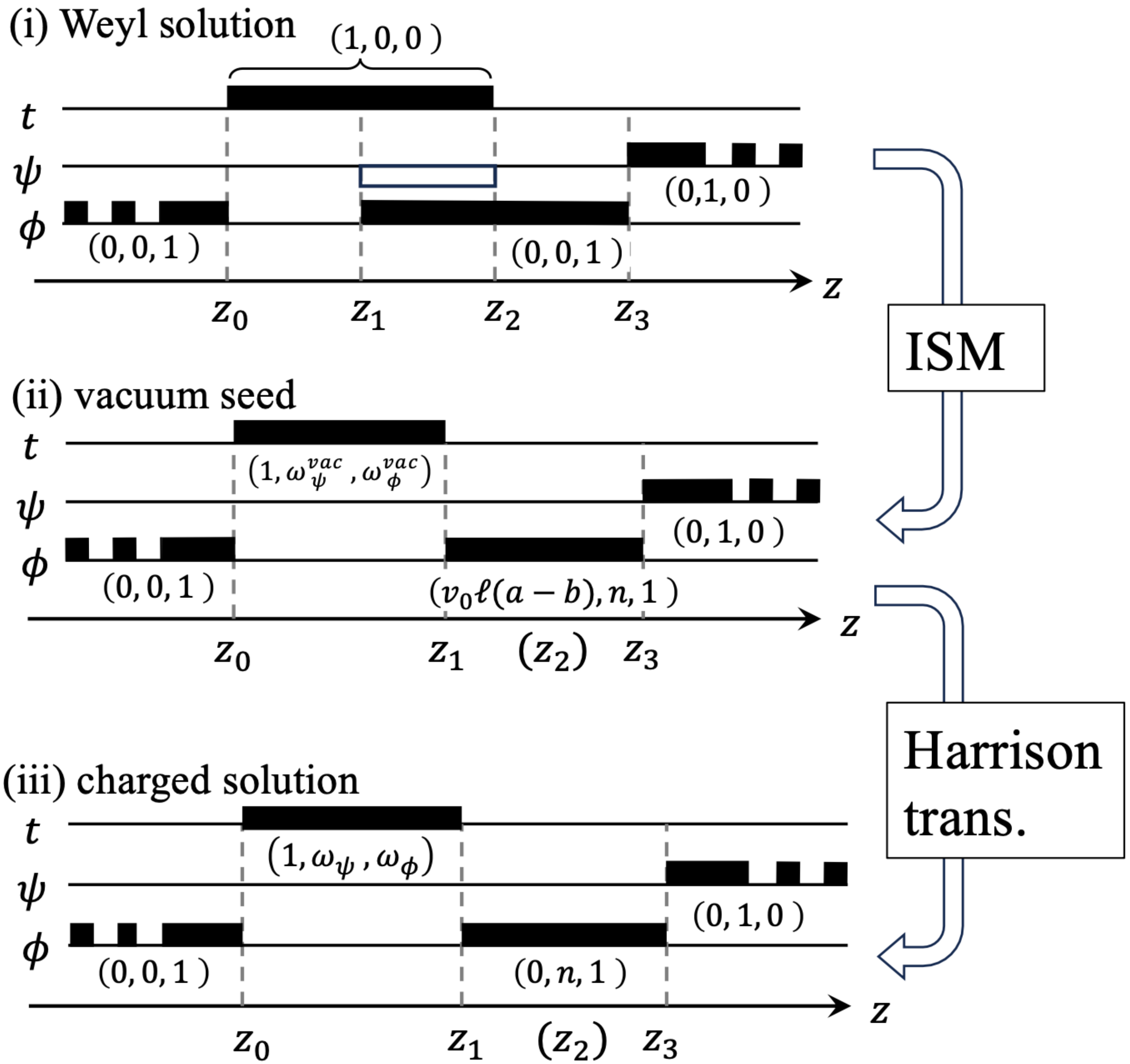}
\caption{The outline of the construction of charged black holes  in Refs.~\cite{Suzuki:2024vzq,Suzuki:2023nqf,Suzuki:2024phv}. The ISM transforms the rod structure from (i) to (ii), where (i) represents a Weyl solution described by a diagonal metric, and (ii), the vacuum seed solution with the Dirac-Misner string singularity on the rod $\{ (\rho,z) \mid \rho=0, z_1<z<z_3]\}$ due to the time component in the rod vector.
This singularity is removed by the Harrison transformation with the finely tuned parameter, leading to the rod structure of (iii). (iii) corresponds to the metric solution for charged black rings of $S^2 \times S^1$ topology for $n=0$ and   black holes of $S^3$-topology having a $2$-cycle on the rod $\{ (\rho,z) \mid \rho=0, z_1<z<z_3]\}$  (capped black holes) for $n=\pm1$.
 \label{fig:harrison-prev}}
\end{figure}

 \medskip
However, in all known cases, the Harrison transformation produces a type of singularity called the Dirac-Misner string singularity, when it is applied to the spacetime 
having 2-cycles such as black rings and capped black holes~\cite{Bouchareb:2007ax,Suzuki:2024vzq,Suzuki:2023nqf,Suzuki:2024phv,Suzuki:2024coe,Suzuki:2024abu}.
A resolution to this issue is applying the transformation to the vacuum solution possessing the Dirac-Misner string singularity so that the singularity can be removed by tuning the parameter of the transformation~\cite{Suzuki:2024vzq,Suzuki:2023nqf,Suzuki:2024phv,Suzuki:2024coe,Suzuki:2024abu} (Fig.~\ref{fig:harrison-prev}).
We follow the same strategy, i.e. (i) construct the vacuum seed with the Dirac-Misner string singularities using the inverse scattering method (ISM),
(ii) apply the Harrison transformation to the vacuum seed and tune the charge parameter so that the Dirac-Misner string singularities are removed.
Then, we determine the conditions for the absence of conical and orbifold singularities and also for the metric remaining to be Lorentzian, respectively.
Finally, we search for the parameter region satisfying these conditions. 

\medskip

The rest of the paper is organized as follows.
In Sec.~\ref{sec:vac}, we construct the vacuum seed metric using the ISM.
In Sec.~\ref{sec:har}, we present the boundary conditions for the charged solution obtained by applying the Harrison transformation to the vacuum seed.
The existence of the solution satisfying the boundary conditions is studied in Sec.~\ref{sec:sols}.
Finally, we summarize our result in Sec.~\ref{sec:sum}.

\section{Vacuum seed}\label{sec:vac}

\medskip

 In this section, 
 following the procedure in the previous works~\cite{Suzuki:2024coe,Suzuki:2023nqf,Suzuki:2024phv}(Fig.~\ref{fig:harrison-prev}), 
 we use the ISM~\cite{Belinsky:1979mh,Belinski:2001ph,Pomeransky:2005sj} to construct the vacuum seed solution, which is later used to produce the charged solution with two bubbles adjacent to the horizon through the Harrison transformation~(Fig.~\ref{fig:rod-sol}).

\subsection{Construction of the vacuum seed}

\begin{figure}[t]
\includegraphics[width=9cm]{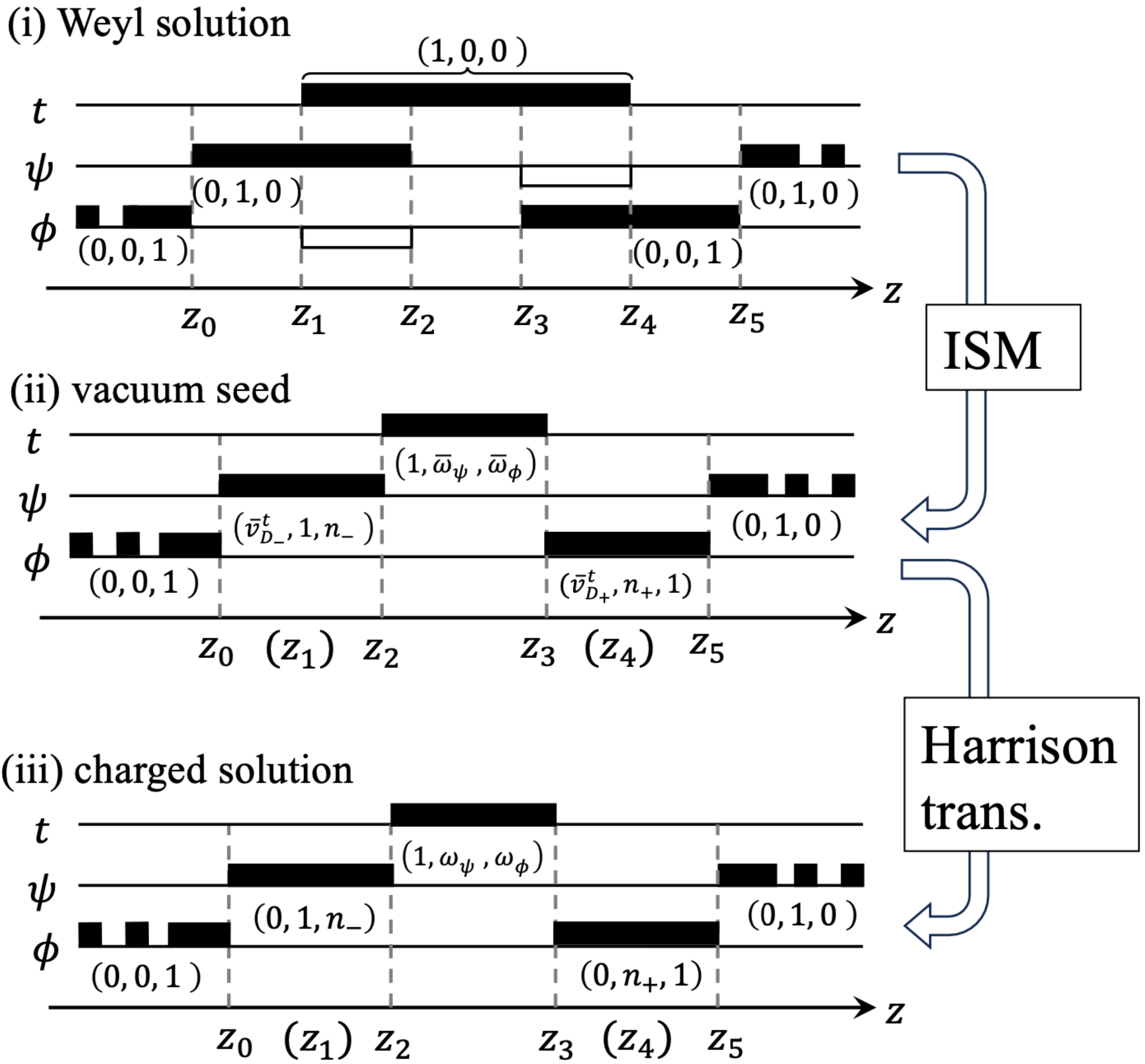}
\caption{The outline of the construction of charged black holes with two bubbles following the procedure  in Refs.~\cite{Suzuki:2024coe,Suzuki:2023nqf,Suzuki:2024phv}. The ISM transforms the rod structure from (i) to (ii), where (i) represents a Weyl solution described by a diagonal metric, and (ii), the vacuum seed solution with the Dirac-Misner string singularity on the rods $\{ (\rho,z) \mid \rho=0, z_0<z<z_2]\}$ and $\{ (\rho,z) \mid \rho=0, z_3<z<z_5]\}$ due to the time component in the rod vectors.
This singularity is removed by the Harrison transformation with the finely tuned parameter, leading to the rod structure of (iii). (iii) represents the rod structure of the charged black holes with two $2$-cycles on the rods  $\{ (\rho,z) \mid \rho=0, z_0<z<z_2]\}$ and $\{ (\rho,z) \mid \rho=0, z_3<z<z_5]\}$ which  we seek for in this paper.
\label{fig:rod-sol}}
\end{figure}

Assuming three Killing vectors $\xi_0 = \partial_t, \ \xi_1 = \partial_\psi, \ \xi_2= \partial_\phi$,
one can write the spacetime in the Weyl-Papapetrou form as~\cite{Emparan:2001wk}
\begin{align}
  ds^2 = G_{ij}(\rho,z) dx^i dx^j + f(\rho,z) (d\rho^2+dz^2),\label{eq:WPform-vac}
\end{align}
where $(x^i) = (t,\psi,\phi)$ $(i=0,1,2)$. The canonical coordinates $(\rho,z)$ are defined so that the $3\times 3$ matrix $G=(G_{ij})$ follows the constraint ${\rm det}(G)=-\rho^2$, and are globally well-defined, harmonic, and mutually conjugate on a timeslice.
The spacetime on and outside the horizon is given by $(\rho,z) \in [0,\infty) \times (-\infty,\infty)$, where the boundary consists of the $z$-axis ($\rho=0$) and the infinity ($\rho^2+z^2 \to \infty$). According to Ref.~\cite{Hollands:2007aj}, provided the spacetime is asymptotically flat and does not have orbifold, conical, Dirac-Misner string and curvature singularities on and outside the horizon, the geometry given by Eq.~(\ref{eq:WPform-vac}) is uniquely determined by the rod structure~\cite{Harmark:2004rm} at $\rho=0$ and the asymptotic charges at the infinity (mass and angular momenta)
, and hence we focus on the regularity of the boundary geometry.

\medskip

 We begin with the following Weyl solution (Fig~\ref{fig:rod-sol})
\begin{align}
&G_0 = {\rm diag} \left\{-\frac{\mu_0}{\mu_4} , \frac{\mu_0 \mu_4 \mu_5}{\mu_2 \mu_3}, \frac{\rho^2 \mu_2 \mu_3}{\mu_0 \mu_1\mu_5} \right\},\\
& f_0 = \frac{C_f \mu_0 \mu_4 \mu_5 R_{02}^2 R_{03}^2 R_{12} R_{13} R_{14} R_{24} R_{25}^2R_{34}R_{35}^2}{\mu_2\mu_3  R_{00} R_{01} R_{04} R_{05}^2 R_{11} R_{15}  R_{22}R_{23}^2 R_{33} R_{44} R_{45} R_{55}},
\end{align}
where $\mu_i := \sqrt{\rho^2+(z-z_i)^2}-z+z_i$ and $R_{ij} := \rho^2+\mu_i \mu_j$.

\medskip
Then, we perform the soliton transformation of this metric as follows.
According to Pomeransky's instruction~\cite{Pomeransky:2005sj}, we first remove
\begin{itemize}
\item an anti-soliton with the trivial Belisky-Zakharov (BZ) vector $(1,0,0)$  at $z=z_1$,
\item a soliton with the trivial BZ vector $(0,1,0)$ at $z=z_2$, 
\item an-anti soliton with the trivial BZ vector $(0,0,1)$ at $z=z_3$,
\item a soliton with the trivial BZ vector $(1,0,0)$ at $z=z_4$.
\end{itemize}
This operation simply multiplies each diagonal component of $G_0$ by corresponding factors 
\begin{align}
\tilde{G}_0 = {\rm diag}\left\{\left(-\frac{\rho^2}{\mu_1^2}\right)\left(-\frac{\mu_4^2}{\rho^2}\right) ,-\frac{\mu_2^2}{\rho^2},
-\frac{\rho^2}{\mu_3^2}\right\} G_0.
\end{align}
Next, we readd to $\tilde{G}_0$
\begin{itemize}
\item  an anti-soliton with the BZ vector $m_{0,1}=(1,0,C_1)$ at  $z=z_1$,
\item a soliton with the BZ vector $m_{0,0}=(0,1,C_0)$  at $z=z_2$,
\item an anti-soliton with the BZ vector  $m_{0,3}=(0,C_3,1)$ at  $z=z_3$,
\item a soliton with the BZ vector $m_{0,4}=(1,C_4,0)$ at $z=z_4$.
\end{itemize}
This leads to the new metric solution expressed by
\begin{align}
\bar{G}_4 = \tilde{G}_0 - \sum_{i,j=1,2,3,4}\frac{(\Gamma^{-1})_{ij} (\tilde{G}_0 m_i) \otimes (\tilde{G}_0 m_j)}{\mu_i \mu_j},\quad \Gamma_{ij} :=  \frac{m_i^T \tilde{G}_0 m_j}{\rho^2+\mu_i\mu_j},\quad m_i := \Psi_0(\lambda=\mu_i,\rho,z)^{-1} m_{0,i},\label{eq:sol-g4}
\end{align}
with
\begin{align}
f_4 = \frac{{\rm det} G_4}{{\rm det} G_4\bigr|_{C_i=0}}f_0.
\label{eq:sol-f4}
\end{align}
The generating matrix $\Psi_0(\lambda,\rho,z)$ is obtained by the replacement $\rho^2 \to \rho^2-2 \lambda z-\lambda^2,\ \mu_i \to \mu_i-\lambda, \bar{\mu}_i \to \bar{\mu}_i-\lambda$ in
\begin{align}
&\tilde{G}_0 = {\rm diag}\left\{\frac{\rho ^2}{\mu _1 \bar{\mu} _4},-\frac{\mu _0 \mu _5 \rho ^2}{\mu _3  \bar{\mu}_2 \nu _4},\frac{\rho ^6}{\mu _0 \mu _1 \mu _3 \mu _5  \bar{\mu} _2}\right\}
\end{align}
where $\bar{\mu}_i = -\sqrt{\rho^2+(z-z_i)^2}-z+z_i$ and we used the identity $\mu_i \bar{\mu}_i = -\rho^2$.

\medskip
Below, we express the solution with the following parameters instead of $C_i$
\begin{align}
C_1= \frac{a z_1^2 \sqrt{2z_{41}}}{z_{21}z_{31}},\quad
C_2 = \frac{p z_2^2 z_{32} z_{42} }{ z_{20}^2 z_{52}^2},\quad C_3 = \frac{q z_{31}z_{32}z_{43}}{z_3^2 z_{43} },\quad
 C_4 = -\frac{b z_{40} z_{54} \sqrt{2z_{41} }}{z_{42}z_{43}},
   \end{align}
   where $z_{ij}:=z_i-z_j$.
It is worth noting that, due to the symmetric setup, the solution~(\ref{eq:sol-g4}) and (\ref{eq:sol-f4}) is invariant under the following transformation
\begin{align}
z\to -z, \quad \psi \leftrightarrow \phi, \quad z_i \to - z_{5-i},\quad a \leftrightarrow b,\quad p\leftrightarrow q,\label{eq:symmetry}
\end{align}
leading to
\begin{align}
 \mu_i \to - \bar{\mu}_{5-i},\quad R_{ij} \to  \frac{\rho^2 R_{5-j\ 5-i}}{\mu_{5-i} \mu_{5-j}}.
\end{align}

\medskip

At the infinity $\sqrt{\rho^2+z^2}\to\infty$, the new solution $\bar{G}_4$ and $f_4$ describes the asymptotically flat spacetime in the rotating frame.
The asymptotic form becomes the Minkowski metric in the rest frame via the transformation
\begin{align}
G_4 = \Lambda^{T} \bar{G}_4 \Lambda
\end{align}
where
\begin{align}
\Lambda := \left(
\begin{array}{ccc}
1& -\beta v_0/(2bz_{42}) &-\beta v_0 /(2a z_{31})\\
0&v_0&\displaystyle -\beta v_0\\
0&\displaystyle  -\beta v_0 &v_0
\end{array}\right),\quad \beta:=\frac{a b}{1+a b (p+q)},\quad 
v_0 = \fr{\sqrt{1-\beta^2}}.\label{eq:def-globalrot}
\end{align}
Note that this transformation is possible only if $-1<\beta<1$.

The points $(\rho,z)=(0,z_1)$ and $(\rho,z)=(0,z_4)$ become curvature singularities unless the following constraints are imposed
\begin{align}
a^2 = \frac{z_{21}z_{31}}{z_{10}z_{51}},\quad
b^2 = \frac{z_{42}z_{43}}{z_{40}z_{54}}.\label{eq:remnegrod}
\end{align}
Under these constraints, 
the two rods $\{(\rho,z) \mid \rho=0, z \in [z_0,z_1]\}$ and $\{(\rho,z) \mid \rho=0, z \in [z_1,z_2]\}$($\{(\rho,z) \mid \rho=0, z \in [z_3,z_4]\}$ and $\{(\rho,z) \mid \rho=0, z \in [z_4,z_5]\}$) merge into a single rod $\{(\rho,z) \mid \rho=0, z \in [z_0,z_2]\}$ ($\{(\rho,z) \mid \rho=0, z \in [z_3,z_5]\}$).

\medskip

After removing the rotation at the infinity and imposing Eq.~(\ref{eq:remnegrod}), the vacuum seed is characterized by the boundary geometry that consists of the asymptotic infinity $I_\infty = \{(\rho,z)\ |\ \rho^2+z^2\to \infty, |z|/\sqrt{\rho^2+z^2}<\infty\}$ and the rod structure at $\rho=0$.

\subsection{Asymptotic infinity}
In the spherical coordinates $(r,\theta)$
\begin{align}
 \rho = \frac{r^2}{2} \sin(2\theta),\quad
  z = \frac{r^2}{2} \cos(2\theta),\label{eq:coord-sph}
\end{align}
one can see that the metric approaches the standard metric of the five-dimensional Minkowski spacetime at $r\to \infty$
\begin{align}
ds^2 \simeq -dt^2 +r^2 \sin^2\theta d\psi^2+r^2 \cos^2\theta d\phi^2 + dr^2 + r^2 d\theta^2,
\end{align}
where we set the scaling constant as
\begin{align}
C_f := \frac{\beta^2}{(1-\beta^2)a^2b^2}.
\end{align}
Then, the vacuum seed solution is asymptotically flat.

\subsection{Rod structure}

\begin{enumerate}

\item{The $\phi$-rotational axis: $I_{-} = \{(\rho,z)\ |\  \rho=0, z\in (-\infty,z_0]\}$ with the rod vector $v_- = (0,0,1)$, which is free from the conical singularity with the periodicity of $2\pi$,}
\item{The rotational axis: $I_{D_-} = \{(\rho,z)\ |\  \rho=0, z\in [z_0,z_2]\}$} with the rod vector $v_{D_-} = (\bar{v}_{D_-}^t,1,\bar{v}_{D_-}^\phi)$, 
where
\begin{align}
&\bar{v}^t_{D_-} :=\frac{  \sqrt{2z_{41}} \left(b z_{20} z_{30} \left(a^2 (1-2 p q)+1\right)+a q z_{20} \left(b
   z_{40} (a p+b)+z_{43} (a b q+1)\right)+a p z_{30} \left(b z_{40} (a q+b)+z_{42} (a b
   p+1)\right)\right)}{\sqrt{1-\beta^2}(1+a b (p+q)) \left(a b z_{40} \left(q z_{20}+p z_{30}\right)+z_{20}
   z_{30}\right)}  ,\\
&\bar{v}^\phi_{D_-}:= \frac{1}{\left(a b \left(q z_{10} z_{20}+p z_{10} z_{30}\right)+z_{20} z_{30}\right) \left(a b \left(q
   z_{20} z_{40}+p z_{30} z_{40}\right)+z_{20} z_{30}\right)}\nonum &\quad \times\biggr[ a b \left\{ p q \left(z_{10} \left(z_{20}^2 \left(z_{30}-2 z_{40}\right)+z_{20} z_{30}
   \left(z_{30}+2 z_{40}\right)-2 z_{30}^2 z_{40}\right)+z_{20} z_{30} \left(z_{20} \left(z_{40}-2
   z_{30}\right)+z_{30} z_{40}\right)\right)+z_{20}^2 z_{30}^2 \right\}\nonum
&\quad +a^2 b^2 z_{10} z_{40} \left(2 p q z_{20} z_{30} (p+q)-q z_{20}^2 \left(p^2+p
   q-1\right)-p z_{30}^2 \left(p q+q^2-1\right)\right)\nonum
&\quad   +(abq+1) q z_{20}^2 z_{31} z_{43}+(abp+1)p z_{21} z_{30}^2
   z_{42}\biggr],
\end{align}
with the norm
\begin{align}
\left(\frac{\Delta_-}{2\pi}\right)^2:=\frac{\rho^2 f_4}{||v_{D_-}||^2}\biggr|_{I_{D_-}}= \frac{\left(a b \left(q z_{10} z_{20}+p z_{10} z_{30}\right)+z_{20} z_{30}\right){}^2 \left(a b \left(q
   z_{20} z_{40}+p z_{30} z_{40}\right)+z_{20} z_{30}\right){}^2}{z_{10} z_{20}^2 z_{30}^2 z_{40}
   z_{50}^2 (a b (p+q-1)+1)^2 (a b (p+q+1)+1)^2},\label{eq:norm_vdm}
\end{align}

\item{The event horizon: $I_{\cal H} = \{(\rho,z)\ |\  \rho=0, z\in [z_2,z_3]\}$} with the rod vector $v_{\cal H} = (1,\bar{\omega}_\psi,\bar{\omega}_\phi)$, where
\begin{align}
&\bar{\omega}_\psi :=\frac{ \bar{\kappa}}{z_{30} z_{32} z_{50} z_{52} (a b (p+q)+1) \sqrt{1-\beta^2}\sqrt{z_{31} z_{40} z_{42} z_{51}}} \\
&\times \biggr[a b \left(a p q z_{32}^2 z_{42} z_{51} \left(z_{30} z_{40}-z_{10} z_{43}\right) +b q z_{20} z_{31} z_{40} z_{52}^2 \left(z_{20}-2 z_{30}\right)-a z_{30}^2
   z_{42} z_{51} z_{52}^2\right)\\
&\quad   +b p q z_{21} z_{31} z_{32}^2 z_{40} z_{42} (a b (p+q)+1)+z_{30}^2 (a b p+1) \left(a p z_{32}^2 z_{42} z_{51}-b z_{31} z_{40}
   z_{52}^2\right)\biggr],
\end{align}
where $\bar{\kappa}$ is the surface gravity shown in the Appendix.~\ref{app:kappa}, and
\begin{align}
\bar{\omega}_{\psi} =\bar{\omega}_{\phi}\bigr|_{z_i \to -z_{5-i}, a \leftrightarrow b, p \leftrightarrow q},
\end{align}

\item{The rotational axis: $I_{D_+} = \{(\rho,z)\ |\  \rho=0, z\in [z_3,z_5]\}$}  with the rod vector $v_{D_+} = (\bar{v}_{D_+}^t,\bar{v}_{D_+}^\psi,1)$,
where
\begin{align}
(\bar{v}_{D_+}^t,\bar{v}_{D_+}^\psi) = (\bar{v}_{D_-}^t,\bar{v}_{D_-}^\phi)\biggr|_{z_i \to -z_{5-i}, a \leftrightarrow b, p \leftrightarrow q},
\end{align}
and
\begin{align}
\left(\frac{\Delta_+}{2\pi}\right)^2:=\frac{\rho^2 f_4}{||v_{D_+}||^2}\biggr|_{I_{D_+}}=\frac{\rho^2 f_4}{||v_{D_-}||^2}\biggr|_{I_{D_-},z_i \to -z_{5-i}, a \leftrightarrow b, p \leftrightarrow q}. \label{eq:norm_vdp}
\end{align}
\item{The $\phi$-rotational axis: $I_+ = \{(\rho,z)\ |\  \rho=0, z\in [z_5,\infty)\}$} with the rod vector $v_+ = (0,1,0)$, which is free from the conical singularity with the periodicity of $2\pi$.
\end{enumerate}

\section{Boundary conditions for charged solution}\label{sec:har}

The Harrison transformation~\cite{Bouchareb:2007ax} transforms the vacuum seed metric given by $G_4$ and $f_4$ into the following form
\begin{align}
ds^2 = (G'_4)_{ij} dx^i dx^j + D f_4 (d\rho^2+dz^2),\quad D:=c^2 + s^2 (G_4)_{00},\label{eq:charged-metric}
\end{align}
where $(c,s):=(\cosh\alpha,\sinh\alpha)$. In this paper, we do not present the explicit form of $G_4'$, but instead, focus on its rod structure and the regularity on it. More precisely, we derive the conditions for the absence of orbifold, conical and Dirac-Misner string singularities, as well as an inequality for the spacetime having the Lorentizan signature on the rod. The absence of curvature singularities and CTCs are studied numerically for a certain set of parameters.

\medskip

One can easily check that the spacetime remains to be asymptotically flat at $I_\infty$ by taking the limit $r\to\infty$
with the coordinates~(\ref{eq:coord-sph}). The rest of the boundary behavior changes as follows:

\begin{enumerate}

\item{The  $\phi$-rotational axis: $I_{-} = \{(\rho,z)\ |\  \rho=0, z\in (-\infty,z_0]\}$ with the rod vector $v_- = (0,0,1)$, which is free from the conical singularity with the periodicity of $2\pi$.}
\item{The rotational axis: $I_{D_-} = \{(\rho,z)\ |\  \rho=0, z\in [z_0,z_2]\}$} with the rod vector
$v_{D_-}=(\bar{v}_{D_-}^t(c^3-s^3/\chi_-),1,\bar{v}_{D_-}^\phi)$ where
\begin{align}
&\chi_- = \frac{a
   b z_{10} \left(q z_{20}+p z_{30}\right)+z_{20} z_{30}}{a b z_{40} \left(q z_{20}+p z_{30}\right)+z_{20} z_{30}}\nonum
&\quad \times   \frac{\left(a z_{40} \left(b^2+a b (p+q)+1\right) \left(q z_{20}+p z_{30}\right)+z_{20} z_{30} \left(b \left(a^2 \left(1-(p+q)^2\right)+1\right)-a (p+q)\right)\right) }
{ \left(b z_{10} \left(a^2+a b   (p+q)+1\right) \left(q z_{20}+p z_{30}\right)+z_{20} z_{30} 
\left(a \left(b^2 \left(1-(p+q)^2\right)+1\right)-b (p+q)\right)\right)},
\end{align}
with the norm given by Eq.~(\ref{eq:norm_vdm}).
 
\item{The event horizon: $I_{\cal H} = \{(\rho,z)\ |\  \rho=0, z\in [z_2,z_3]\}$} with the rod vector $v_{\cal H} = (1,\omega_\psi,\omega_\phi)$, where
\begin{align}
\begin{split}
&\omega_\psi :=  \frac{\bar{\omega}_\psi}{c^3+s^3 \bar{\kappa}/\hat{\kappa} },
   \end{split}
\end{align}
with $\hat{\kappa}$ given in the Appendix.~\ref{app:kappa} and
\begin{align}
\omega_{\psi} =\omega_{\phi}\bigr|_{z_i \to -z_{5-i}, a \leftrightarrow b, p \leftrightarrow q}.
\end{align}
The surface gravity is given by
\begin{align}
\kappa = (c^3\bar{\kappa}^{-1}+s^3 \hat{\kappa}^{-1})^{-1},
\end{align}

\item{The rotational axis: $I_{D_+} = \{(\rho,z)\ |\  \rho=0, z\in [z_3,z_5]\}$} with the rod vector $v_{D_+} = (\bar{v}_{D_+}^t(c^3-s^3/\chi_+),\bar{v}_{D_+}^\psi,1)$,
where
\begin{align}
\chi_+ = \chi_-\bigr|_{z_i \to -z_{5-i}, a \leftrightarrow b, p \leftrightarrow q},
\end{align}
with the norm given by Eq.~(\ref{eq:norm_vdp}).

\item{The $\phi$-rotational axis: $I_+ = \{(\rho,z)\ |\  \rho=0, z\in (z_5,\infty)\}$ with the rod vector $v_+ = (0,1,0)$, which is free from the conical singularity with the periodicity of $2\pi$.}
\end{enumerate}

\medskip

\subsection{Boundary conditions}
To obtain the physical spacetime metric, we impose the conditions below on the above boundary data.

\medskip
\paragraph{Spacetime signatures at $z=z_0$ and $z=z_5$}

Let us consider the expansion around $z=z_0$.
By taking the limit $r\to 0$ with $(\rho,z-z_0) = |{\cal C}_-|^{-1} r^2 (\sin(2\theta),\cos(2\theta))$, 
we find that the metric behaves as
\begin{align}
ds^2 \simeq - dt'{}^2 + \frac{|{\cal C}_-|}{{\cal C}_-} \left(dr^2 + r^2 d\theta^2 + \left(\frac{2\pi}{\Delta_-}\right)^2 r^2 \sin^2\theta d\phi_1^2 + r^2 \cos^2 \theta d\phi_2{}^2 \right),\label{eq:metric-z0}
\end{align}
where $t':= {\cal A}_- (t-v_{D_-}^t \psi)$, $\phi_1:=\psi,\ \phi_2:=\phi- v^\phi_{D_-} \psi$ and
\begin{align}
&{\cal C}_- = \frac{2  \left(c^2 z_{40}\left(z_{20}
   z_{30}+a b z_{10} \left(q z_{20}+p z_{30}\right)\right){}^2-s^2 z_{10}\left( z_{20} z_{30}+a bz_{40} (q z_{20}+ p z_{30} ) \right){}^2\right)}{(1-\beta^2)z_{10}z_{20} z_{30}z_{40} z_{50} (a b (p+q)+1)^2},\\
&{\cal A}_- = \frac{2 \left(a b \left(q z_{10} z_{20}+p z_{10} z_{30}\right)+z_{20} z_{30}\right) \left(a b
   \left(q z_{20} z_{40}+p z_{30} z_{40}\right)+z_{20} z_{30}\right)}{(1-\beta^2)z_{20} z_{30} z_{50} \sqrt{z_{10}
   z_{40}} (a b (p+q)+1)^2 {\cal C}_-}.
\end{align}
This spacetime possesses the conical singularity if $\Delta_-$ given in Eq.(\ref{eq:norm_vdm}) is not $2\pi$. Regardless of the existence of the conical singularity, the metric becomes  Lorentzian if and only if
\begin{align}
{\cal C}_->0 \Longleftrightarrow \tanh^2\alpha < (t^-_{max})^2 := \frac{z_{40}(z_{20}z_{30}+ab z_{10}(qz_{20}+p z_{30}))^2}{z_{10}(z_{20}z_{30}+abz_{40}(q z_{20}+pz_{30}))^2}.\label{eq:t2bound-m}
\end{align}
The expansion around $z=z_5$ is obtained from the above results by the replacement
in Eq.~(\ref{eq:symmetry}) along with
\begin{align}
{\cal C}_- \to {\cal C}_+ := {\cal C}_- \biggr|_{z_i\to -z_{5-i},a\leftrightarrow b,p\leftrightarrow q},\quad 
{\cal A}_- \to {\cal A}_+ := {\cal A}_-  \biggr|_{z_i\to -z_{5-i},a\leftrightarrow b,p\leftrightarrow q},\quad
\Delta_- \to \Delta_+,
\end{align}
whose signature becomes Lorentizan if and only if
\begin{align}
{\cal C}_+ >0 \Longleftrightarrow \tanh^2\alpha < (t^+_{max})^2 := (t^-_{max})^2\biggr|_{z_i \to -z_{5-i},a\leftrightarrow b, p \leftrightarrow q}.\label{eq:t2bound-p}
\end{align}

\paragraph{Absence of orbifold singularity}
According to the discussion in Ref.~\cite{Harmark:2004rm}, 
the spacetime is free from the orbifold singularity if
\begin{align}
\begin{split}
 \bar{v}^\phi_{D_-}=n_-,\quad \bar{v}^\psi_{D_+} =n_+,\quad n_\pm \in {\mathbb Z},\label{eq:topcon-sym-0}
 \end{split}
\end{align}
leading to the horizon topology of $S^3$ for $1-n_- n_+=\pm 1$, $S^2 \times S^1$ for $1-n_- n_+= 0$ and $L(|1-n_- n_+|;1)$ for $1-n_- n_+=\pm 2,\pm 3,\dots$.

\medskip

\paragraph{Absence of Dirac-Misner string singularity}

The Dirac-Misner string singularity is absent if the inner axes do not have the time component in the rod vectors
   \begin{align}
 v^t_{D_\pm} = \bar{v}^t_{D_\pm} (c^3-s^3/\chi_\pm) = 0  \quad
\Longleftrightarrow \quad \tanh^3 \alpha =\chi_\pm,\label{eq:nodms-0}
   \end{align}
meaning that   the Dirac-Misner string can be removed if and only if the following condition holds
\begin{align}
|\chi_\pm|<1,\quad \chi_+ = \chi_-.\label{eq:nodms-cond-0}
\end{align}

\medskip

\paragraph{Absence of conical singularity }   

Once the Dirac-Misner string singularity is removed after the transformation, $I_{D_-}$ ($I_{D_+}$) becomes the regular $\psi$-rotational axis  ($\phi$-rotational axis) with the periodicity $2\pi$ if
\begin{align}
\left(\frac{\Delta_-}{2\pi}\right)^2=1=\left(\frac{\Delta_+}{2\pi}\right)^2.
\label{eq:psi-period-sym-0}
\end{align}

\medskip

\subsection{Boundary conditions in the symmetric case}

In this setup, at most, we have ten parameters 
$\{\alpha, \ z_{i+1\, i} \, (i=0,...,4), \ C_j\, ( j=1,2,3,4)\}$
 constrained by eight equations [Eqs.~(\ref{eq:remnegrod}), (\ref{eq:topcon-sym-0}), (\ref{eq:nodms-0}),  and  (\ref{eq:psi-period-sym-0})].
Because it is difficult to explore the entire parameter space, instead,
 we particularly focus on the spacetime symmetric under the exchange of $\psi$ and $\phi$ as well as $z\to -z$, which halves the numbers of parameters and equations.
For the sake of the symmetry~(\ref{eq:symmetry}),
this can be accomplished by setting
\begin{align}
 b = a ,\quad q = p, \quad z_{5}=-z_0 = \ell^2,\quad z_{4}=-z_1 = \gamma \ell^2,\quad z_{3} = -z_2 = \nu \ell^2,
 \label{eq:setp-sym}
\end{align}
where the parameters for the rod positions run in the range of
\begin{align}
\ell>0,\quad 0<\nu<\gamma<1. \label{eq:psym-range}
\end{align}
The smoothness condition~(\ref{eq:remnegrod})  becomes
\begin{align}
a = \sqrt{\frac{\gamma^2-\nu^2}{1-\gamma^2}}.
 \label{eq:seta-sym}
\end{align}
Using Eqs.~(\ref{eq:seta-sym}), one can show $\beta$ in Eq.~(\ref{eq:def-globalrot}) is expressed by
\begin{align}
\beta =\frac{\gamma ^2-\nu ^2}{2(\gamma ^2 -\nu^2)p+\gamma^2-1},
\end{align}
or in turn, $p$ is expressed by $\beta$ as
\begin{align}
p = \frac{\beta(1-\gamma^2)+\gamma^2-\nu^2}{2\beta(\gamma ^2-\nu^2)}.\label{eq:setp-sym}
\end{align}

From Eqs.~(\ref{eq:seta-sym}) and (\ref{eq:setp-sym}), the symmetric solution is now expressed by five parameters $(\ell,\nu,\gamma,\beta,\alpha)$,
 and the boundary conditions above are expressed as follows:

\medskip

\paragraph{Lorentzian signature around $z=z_0$ and $z=z_5$}
The inequalities~(\ref{eq:t2bound-m}) and (\ref{eq:t2bound-p}) reduce to
\begin{align}
 \tanh^2\alpha < t_{max}^2 
:= \frac{(1-\gamma ) \left(\beta  \left(\gamma ^2-\nu ^2+\gamma  (1-\nu ^2)\right)+\gamma ^2-\nu ^2\right)^2}{(1+\gamma ) \left(\beta  \left(\gamma ^2-\nu ^2-\gamma  \left(1-\nu   ^2\right)\right)+\gamma ^2-\nu ^2\right)^2}.\label{eq:t2bound}
\end{align}

\paragraph{Absence of Dirac-Misner string singularity}
The condition~(\ref{eq:topcon-sym-0}) becomes
\begin{align}
\begin{split}
&\bar{v}^\phi_{D_-} =\bar{v}^\psi_{D_+} =\frac{1}{\beta  \left(\beta  \left(\gamma ^2-\gamma  \left(1-\nu ^2\right)-\nu ^2\right)+\gamma ^2-\nu ^2\right) \left(\beta  \left(\gamma ^2+\gamma 
   \left(1-\nu ^2\right)-\nu ^2\right)+\gamma ^2-\nu ^2\right)}\\
&\quad \times   \biggr\{ \left(\gamma ^2-\nu ^2\right)\left[\beta ^3 \left(\gamma ^2 \left(\nu ^2+1\right)-\nu ^2 \left(3-\nu ^2\right)\right)+\beta  \left(1-\nu ^2\right) \left(\gamma
   ^2+\nu ^2\right)-\nu ^2 \left(\gamma ^2-\nu ^2\right)\right]\\
&\quad\quad    +\beta ^2 \left(\left(\nu ^2+2\right) \left(\gamma ^4+\nu ^4\right)-\gamma ^2 \left(3 \nu ^4+2 \nu
   ^2+1\right)\right)\biggr\}=n \in {\mathbb Z},\label{eq:topcon-sym}
 \end{split}
\end{align}
where the horizon topology becomes $S^3$ for $n=0$, $S^2 \times S^1$ for $n=\pm1$ and $L(n^2-1,1)$ for $n=\pm 2,\pm 3,\dots$.

\medskip

\paragraph{Absence of conical singularity }   
The condtion~(\ref{eq:nodms-0}) reduces to
 \begin{align}
&\tanh^3 \alpha =\frac{(1-\gamma ) \left(\beta  (\gamma +1) \left(\gamma -\nu ^2\right)+\gamma ^2-\nu ^2\right) \left(\beta  \left(\gamma ^3+\gamma ^2-\gamma +\nu ^4-2 \nu
   ^2\right)+\left(\gamma ^2-\nu ^2\right) \left(\gamma +\nu ^2\right)\right)}{(\gamma +1) \left(\beta  \left(\gamma ^3-\gamma ^2-\gamma -\nu ^4+2 \nu ^2\right)+\left(\gamma
   -\nu ^2\right) \left(\gamma ^2-\nu ^2\right)\right) \left(\beta  \left(\gamma ^2+\gamma  \left(\nu ^2-1\right)-\nu ^2\right)+\gamma ^2-\nu ^2\right)},\label{eq:nodms}
   \end{align}
and the inequality~(\ref{eq:nodms-cond}) becomes
\begin{align}
\left|\frac{(1-\gamma ) \left(\beta  (\gamma +1) \left(\gamma -\nu ^2\right)+\gamma ^2-\nu ^2\right) \left(\beta  \left(\gamma ^3+\gamma ^2-\gamma +\nu ^4-2 \nu
   ^2\right)+\left(\gamma ^2-\nu ^2\right) \left(\gamma +\nu ^2\right)\right)}{(\gamma +1) \left(\beta  \left(\gamma ^3-\gamma ^2-\gamma -\nu ^4+2 \nu ^2\right)+\left(\gamma
   -\nu ^2\right) \left(\gamma ^2-\nu ^2\right)\right) \left(\beta  \left(\gamma ^2+\gamma  \left(\nu ^2-1\right)-\nu ^2\right)+\gamma ^2-\nu ^2\right)}\right| <1.\label{eq:nodms-cond}
\end{align}

\medskip

\paragraph{Absence of conical singularity }  
The condition~(\ref{eq:psi-period-sym-0}) reduces to
   \begin{align}
\left(\frac{\Delta_\pm}{2\pi} \right)^2=\frac{(1-\gamma^2 )  \left(\beta  \left(\gamma ^2-\gamma  \nu ^2+\gamma -\nu ^2\right)+\gamma
   ^2-\nu ^2\right)^2 \left(\beta  \left(\gamma ^2+\gamma  \left(\nu ^2-1\right)-\nu ^2\right)+\gamma
   ^2-\nu ^2\right)^2}{4 (1-\beta^2)^2 (1-\nu ^2)^2 (\gamma^2 -\nu^2 )^4}=1.\label{eq:psi-period-sym-2}
\end{align}

\medskip

Hence, recalling that the charged symmetric solution possesses five parameters $(\ell,\nu,\gamma,\beta,\alpha)$, the solution satisfying above conditions (\ref{eq:topcon-sym}), (\ref{eq:nodms}) and (\ref{eq:psi-period-sym-2})
is described by two parameters, and has the rod structure illustrated in Fig.~\ref{fig:rod-sol}.

\medskip

\section{On the existence of charged solutions satisfying boundary conditions}
\label{sec:sols}

Now, we search for the charged solution satisfying boundary conditions above.
For this, we solve Eqs.~(\ref{eq:topcon-sym}), (\ref{eq:nodms}) and (\ref{eq:psi-period-sym-2}) under the assumptions of Eqs.~(\ref{eq:t2bound}) and (\ref{eq:nodms-cond}). 
In fact, it turns out that the solution satisfying all conditions~(\ref{eq:topcon-sym}), (\ref{eq:nodms}) and (\ref{eq:psi-period-sym-2}) does not exist.
This can be shown as follows.
From Eqs.~(\ref{eq:topcon-sym}) and (\ref{eq:psi-period-sym-2}), one express the square of $\gamma$ in terms of $\nu$ and $\beta$ as
\begin{align}
\gamma^2 = -\frac{(\beta +1) \nu ^4 \left({\cal G}_6 \beta ^6
+ {\cal G}_5\beta^5
   +{\cal G}_4 \beta^4
   + {\cal G}_3\beta^3
   + {\cal G}_2\beta^2
   +  {\cal G}_1\beta +{\cal G}_0 \right)}
   { {\cal H}_7\beta^7+  {\cal H}_6\beta^6 +  {\cal H}_5\beta ^5+
     {\cal H}_4 \beta ^4+  {\cal H}_3 \beta ^3+  {\cal H}_2\beta ^2+
     {\cal H}_1 \beta+ {\cal H}_0},\label{eq:defgamsym}
\end{align}
where ${\cal G}_i$ and ${\cal H}_i$ are polynomials of $n$ and $\nu$ shown in Appendix.~\ref{app:cfs}.

Eliminating $\gamma$ in Eqs.~(\ref{eq:topcon-sym}) and (\ref{eq:psi-period-sym-2}), we obtain the equation for $\nu$ and $\beta$
\begin{align}
\begin{split}
&\beta ^5 \left[ \nu ^8-60 \nu ^6-66 \nu ^4-12 \nu ^2-16 \left(1-\nu ^2\right) n^5+16 \left(1-\nu ^4\right) n^4+4 \left(\nu ^2+1\right) \left(\nu ^2+3\right)^2 n^3\right.\\
&\quad \left.-8 \left(2  \nu ^6+13 \nu ^4+28 \nu ^2+5\right) n^2+32 \left(\nu ^4+7 \nu ^2+4\right) \nu ^2 n+9\right]\\
&+ \beta ^4 \left(-3 \nu ^8+26 \nu ^6-156 \nu ^4-138 \nu ^2+16 \left(\nu ^4-5 \nu ^2+4\right) n^4+32 \left(3 \nu ^4+4 \nu ^2-1\right) n^3\right.\\
&\quad \left.-20 \left(\nu ^6+11 \nu ^4+15 \nu   ^2+5\right) n^2+2 \left(33 \nu ^6+93 \nu ^4+195 \nu ^2+31\right) n+15\right]  \\
&+\beta ^3 \left[2 \left(\nu ^8+37 \nu ^6+23 \nu ^4-81 \nu ^2-12\right)-8 \left(7 \nu ^4-26 \nu ^2+11\right) n^3+8 \left(2 \nu ^6-9 \nu ^4-20 \nu ^2+3\right) n^2\right.\\
&\quad \left.+\left(-71   \nu ^6-59 \nu ^4+251 \nu ^2+71\right) n\right]   \\
&+\beta ^2 \left[2 \nu ^8-33 \nu ^6+119 \nu ^4-3 \nu ^2+8 \left(\nu ^6+24 \nu ^4-23 \nu ^2+6\right) n^2-8 \left(8 \nu ^6+7 \nu ^4-2 \nu ^2+3\right) n-21\right]   \\
&-\beta  \left(1-\nu ^2\right) \left[\left(3-7 \nu ^2\right)^2 n-\left(\nu ^2+3\right)^2 \left(3 \nu ^2+1\right)\right] -\nu ^2 \left(1-\nu ^2\right) \left(\nu ^2+3\right)^2=0.
   \label{eq:conifree-nubeta}
   \end{split}
\end{align}
Because this is a quintic equation with respect $\beta$ and a quantic with respect to $\nu^2$, it is difficult to obtain the explicit solution.
Instead, one can solve numerically Eq.~(\ref{eq:conifree-nubeta}) within the region $0<\nu<1$ and $-1<\beta<1$.
\medskip

\begin{figure}[th]
\includegraphics[width=7cm]{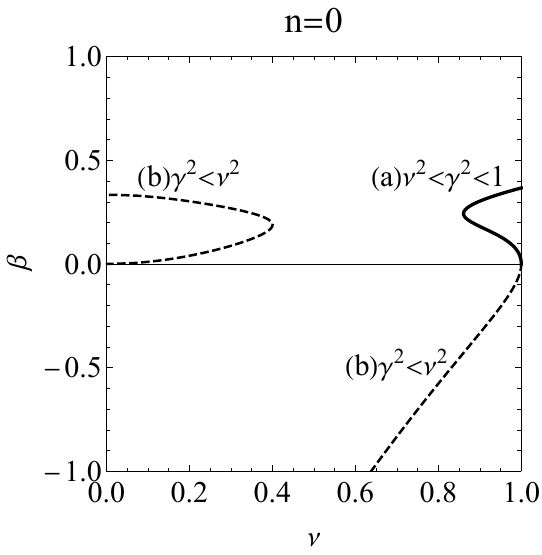}
\caption{The parameter region satisfying Eq.~(\ref{eq:conifree-nubeta}) with $n=0$ in the $(\nu,\beta)$ plane. 
On the thick and dashed curves, Eq.~(\ref{eq:defgamsym}) leads to $\nu^2<\gamma^2<1$ and  $\gamma^2 < \nu^2$, respectively. \label{fig:conitop}}
\end{figure}

\begin{figure}[th]
\includegraphics[width=7cm]{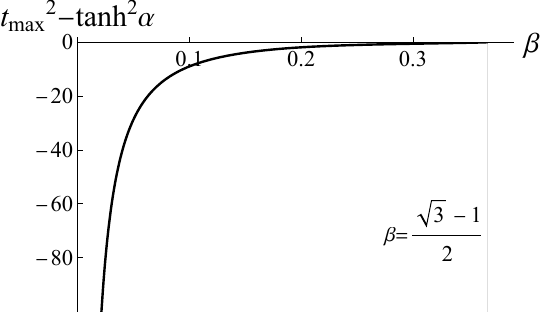}
\caption{Profile of $t_{max}^2-\tanh^2\alpha$ on curve (a) in Fig.~\ref{fig:conitop}. \label{fig:t2bound}}
\end{figure}

\begin{figure}[th]
\begin{minipage}[b]{0.3\columnwidth}
\includegraphics[width=5cm]{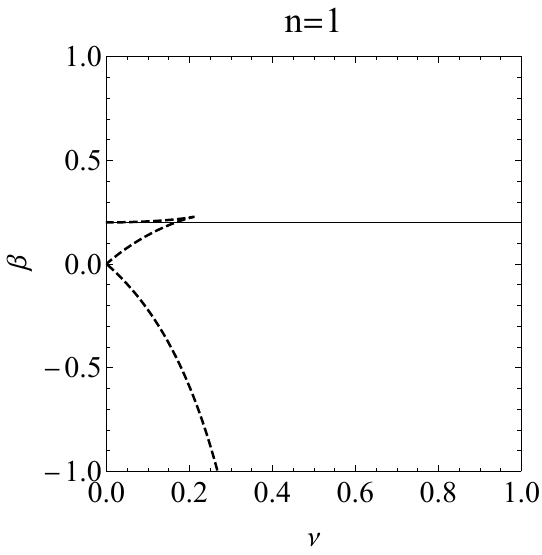}
\includegraphics[width=5cm]{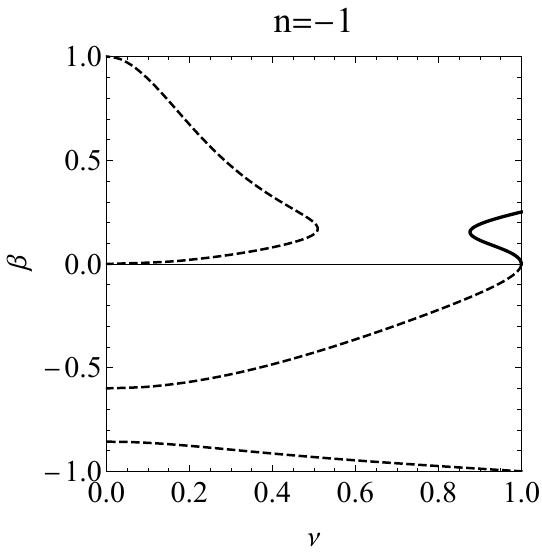}
\end{minipage}
\begin{minipage}[b]{0.3\columnwidth}
\includegraphics[width=5cm]{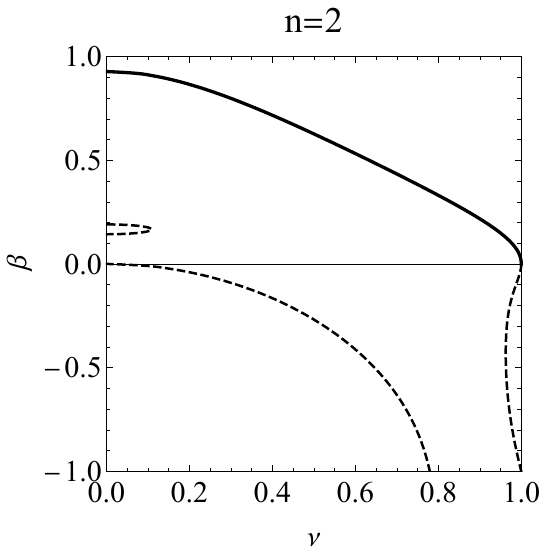}
\includegraphics[width=5cm]{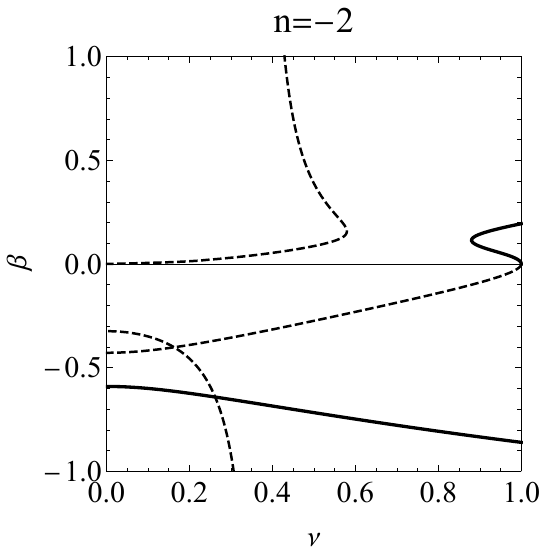}
\end{minipage}
\begin{minipage}[b]{0.3\columnwidth}
\includegraphics[width=5cm]{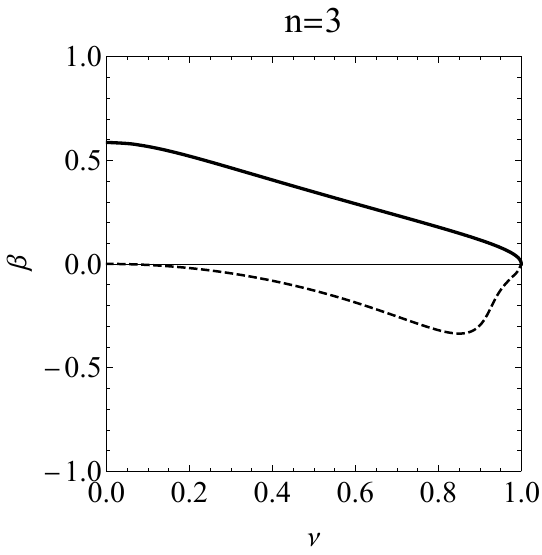}
\includegraphics[width=5cm]{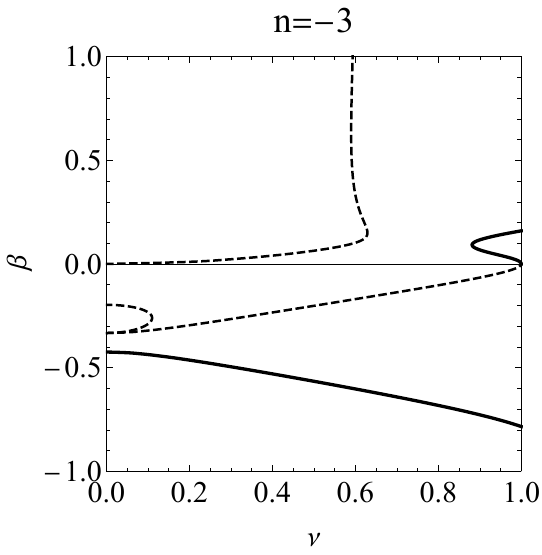}
\end{minipage}
\caption{The parameter region satisfying Eq.~(\ref{eq:conifree-nubeta}) in the $(\nu,\beta)$ plane for the $n=\pm 1,\pm,2,\pm 3$ cases. 
The dashed curves do not exhibit $\gamma$ in the range~(\ref{eq:psym-range}).
The black curves exhibit $\gamma$ within the range~(\ref{eq:psym-range}), but violate the assumption~(\ref{eq:nodms-cond})  in top panels and the assumption~(\ref{eq:t2bound}) in bottom panels.\label{fig:conitopmore}}
\end{figure}

Let us consider, for example, the $n=0$ case where the horizon has $S^3$-topology.
 In Fig.~\ref{fig:conitop}, we present the solution for Eq.~(\ref{eq:conifree-nubeta}) with $n=0$ in the $(\nu,\beta)$-plane.
 We find that Eq.~(\ref{eq:defgamsym}) satisfies the condition~(\ref{eq:psym-range}) only on (a), which exists for $0<\beta<(\sqrt{3}-1)/2$, and (b) is excluded.
 However, the solutions on (a) contradicts the assumption for the Lorentizan metric~(\ref{eq:t2bound})  (Fig.~\ref{fig:t2bound}). Hence, there is no solution for $n=0$.
Repeating the same analysis,
one can confirm that the solution does not exist for $n \neq 0$ as well (Fig.~\ref{fig:conitopmore}).

\medskip

\begin{figure}[th]
\includegraphics[width=5.5cm]{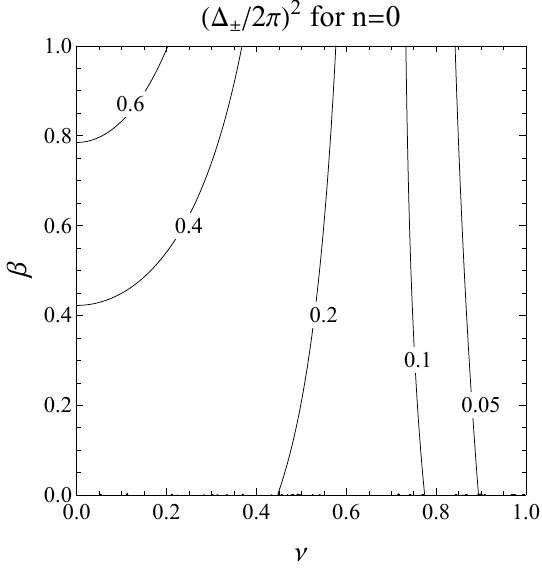}\hspace{0.5cm}
\includegraphics[width=5.5cm]{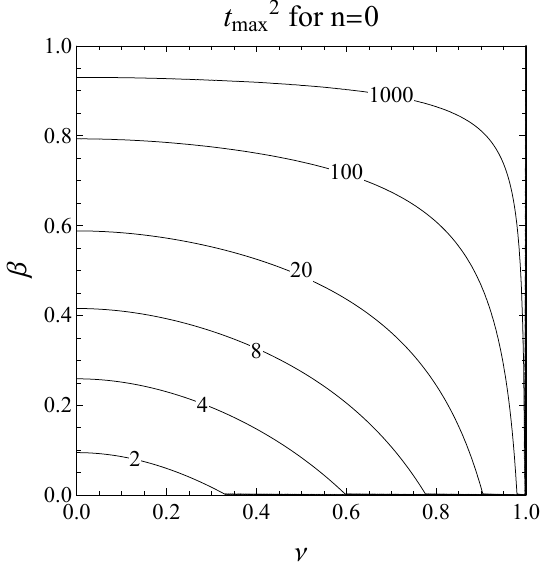}\\
\includegraphics[width=5.5cm]{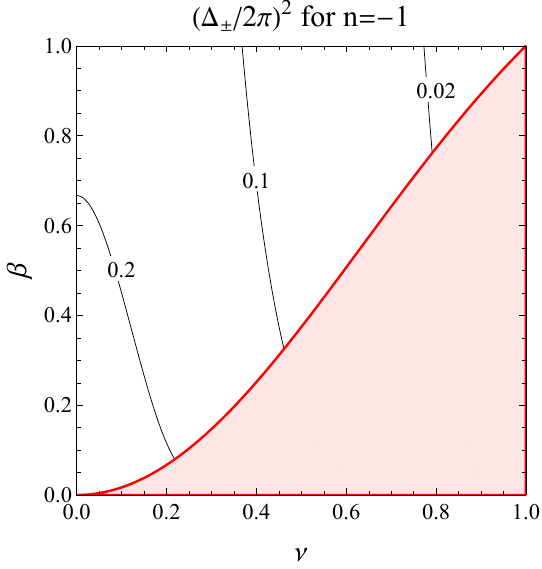}\hspace{0.5cm}
\includegraphics[width=5.5cm]{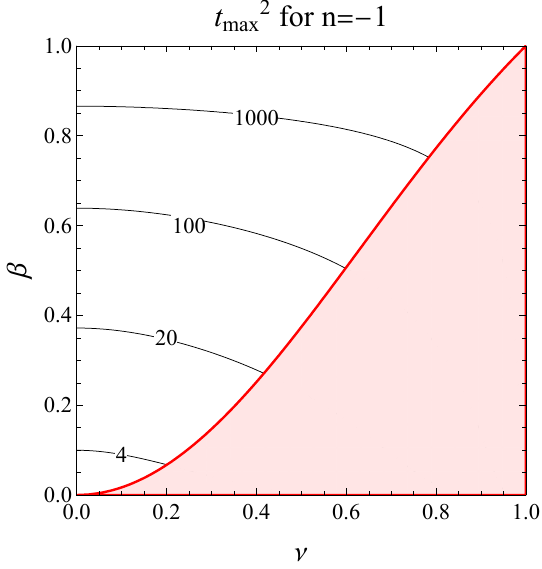}
\caption{Periodicity around $I_{D_\pm}$ and $t_{max}^2$ for the $\gamma=\gamma_+$ solution with $n=0$ ($S^3$-topology) and $n=-1$ ($S^2\times S^1$-topology) in the $(\nu,\beta)$ plane.  In the bottom panels, the red curves correspond to the limit $\tanh\alpha \to -1$ and Eq.~(\ref{eq:nodms-cond}) does not hold in the red colored region.\label{fig:deficit}}
\end{figure}

\begin{figure}[th]
\includegraphics[width=5.5cm]{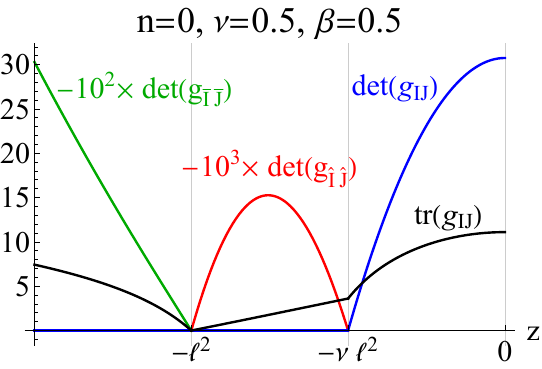}
\hspace{0.5cm}
\includegraphics[width=5.5cm]{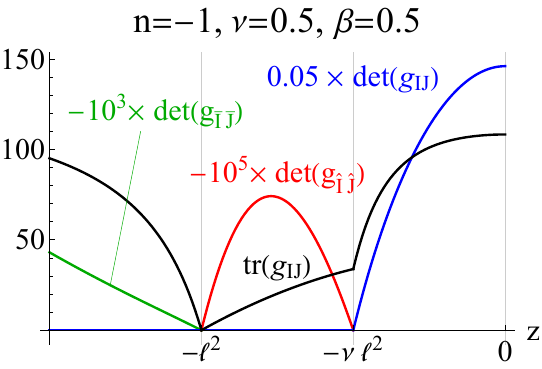}
\caption{
Profiles of ${\rm det}(g_{IJ})$ and ${\rm tr}(g_{IJ})$ on the left half of the rod $\{(\rho,z)\mid \rho=0, z\leq 0\}$, 
${\rm det}(g_{\bar{I}\bar{J}})$ on $I_{-}$, and ${\rm det}(g_{\hat{I}\hat{J}})$ on $I_{D_-}$ for $n=0$ and $n=-1$
 with $(\nu,\beta)=(0.5,0.5)$ and $\gamma=\gamma_+$.\label{fig:noctc}}
\end{figure}

Then, one may ask that, if we allow the conical singularity, can we obtain the solution satisfying all other conditions?
Without the condition~(\ref{eq:psi-period-sym-2}), Eq.~(\ref{eq:topcon-sym}) is solved with respect to $\gamma$ as
\begin{align}
&\gamma_{\pm}^2 = \frac{1}{2 (\beta +1)^2 \left(\nu ^2+\beta 
   \left(n-1-\nu ^2\right)\right)}\biggr[2 \nu ^4
   -\beta ^2 \left(3 \nu ^4+\nu ^2 (2-4 n)+1\right)+\beta ^3
   \left(-4 \nu ^2+\nu ^4 n+n\right)+2 \beta  \nu ^2 n\nonum
&\quad   +\beta ^{3/2} \left(\nu ^2-1\right) \sqrt{\nu ^4 \left(4 \beta ^3-3 \beta +\beta ^3 n^2-6 \beta ^2 n+4 n\right)+2 \nu ^2 (\beta 
   n-1) \left(\left(\beta ^2+4 \beta +2\right) n-\beta  (4 \beta +3)\right)+\beta  (\beta 
   n-1)^2}
   \biggr].
\end{align}
For now, we consider the $n=0$ case.
For the parameter with $\gamma=\gamma_+$, one can show the following properties for the entire range of $(\nu,\beta) \in (0,1)\times (0,1)$:
\begin{itemize}
\item $\gamma_+$ takes a real value with in the range~(\ref{eq:psym-range}),
\item $\Delta_-(=\Delta_+)$ is smaller than $2\pi$ and reaches the maximum of $\sqrt{3}\pi$ at $(\nu,\beta)\to(0,1)$ (left panel in Fig.\ref{fig:deficit}), leading to the conical deficit on $I_{D_\pm}$,
\item The assumption~(\ref{eq:nodms-cond}) holds,
\item $t_{max}^2>1$ for the entire range  (right panel in Fig.~\ref{fig:deficit}) and the assumption~(\ref{eq:t2bound}) holds regardless of the value of $\alpha$.
\end{itemize}
In Fig.~\ref{fig:noctc}, we show the semipositive definiteness and finiteness of $g_{IJ} (I,J=\psi,\phi)$, which is equivalent to $0\leq {\rm det}(g_{IJ})<\infty $ and $0\leq {\rm tr}(g_{IJ}) < \infty$,
exhibiting  the absence of the CTCs on the rod.
In the same figure, we also show  ${\rm det}(g_{\bar{I}\bar{J}})<0\ (\bar{I},\bar{J}=t,\psi)$  on $I_{-}$ and ${\rm det}(g_{\hat{I}\hat{J}})<0 \ (\hat{I},\hat{J}=t,\phi)$ on $I_{D_-}$, assuring that the metric solution  $G_4'$ has a single zero eigenvalue everywhere on the rod except for the end points.
This suggests the absence of the curvature singularities on the rod~\cite{Harmark:2004rm}.
Therefore, the $n=0$ case admits the solution having the conical deficit on the bubbles.
In turn, the parameter with $\gamma=\gamma_-$ never satisfies the assumption~(\ref{eq:t2bound}), and then must be excluded as the solution.

\medskip

We also studied $n=\pm 1,\pm 2, \pm 3$ cases.
As the result, we found that the $n=-1$ case similarly admits the solution possessing conical deficits in a portion of $(\nu,\beta) \in (0,1)\times (0,1)$ (bottom panels in Fig.~\ref{fig:deficit}), except for the red colored region corresponding to the solutions violating Eq.~(\ref{eq:nodms-cond}). One can check the absence of curvature singularities and CTCs for $n=-1$ as well (right panel of Fig.~\ref{fig:noctc}).
For $n=1,\pm 2,\pm 3$,  we have found no solution even in the presence of conical singularities.


\section{Summary} \label{sec:sum}
In this paper, we have investigated the black hole solution with nontrivial DOC topology in five-dimensional supergravity.
For simplicity, we have specifically studied the black hole spacetime having two $2$-cycle structures, or {\it bubbles}, adjacent to the horizon, and also assumed the symmetry between two bubbles.
We have constructed the solution  through the combination of the ISM and Harrison transformation.
We have determined the boundary conditions for the absence of conical, orbifold and Dirac-Misner string singularities as well as the condition for the metric remains Lorentzian on the boundary.
As the result, in the current setup, we have found that the black holes with the horizon topology of $S^3$ and $S^2\times S^1$ exist without any types of singularities except for conical singularities (deficit) on the bubbles.
For other cases such that the horizon topology is $L(3;1)$ or $L(8;1)$, we have found no solution even with conical singularities.

\medskip

In our setup, the conical deficits are inevitable to support the black hole horizon within the bubbles. 
One may interpret this as follows. The conical deficits effectively provide the repulsive force, which prevents the horizon from expanding and swallowing the bubbles. This suggests that the black hole and bubbles can be balanced without conical deficits if we introduce more supporting forces by increasing the magnetic flux.
For other options, one may attach the cosmic strings or expanding horizons at the infinity.

\medskip

The Harrison transformation, in general, produces the Dirac-Misner string singularity on bubbles.
In several cases with a single bubble, it has been shown that this singularity can be eliminated if we transform the vacuum solution possessing it~\cite{Suzuki:2024vzq,Suzuki:2023nqf,Suzuki:2024phv,Suzuki:2024coe,Suzuki:2024abu}. Our result partially shows that the same strategy will be valid for the spacetime with multiple bubbles.
Perhaps, more generic setup might lead to the solution balanced without conical singularities.

\section*{Acknowledgement}
R.S. was supported by JSPS KAKENHI Grant Number JP24K07028. 
S.T. was supported by JSPS KAKENHI Grant Number 21K03560.

\appendix

\section{Explicit forms of some constants}
Below, we present the explicit forms of several quantities used in the main part.
\subsection{Surface gravity}\label{app:kappa}
\begin{align}
\begin{split}
&\bar{\kappa}^{-1} = \frac{ v_0^2 \sqrt{2z_{41}} }{(a b (p+q)+1)^2 z_{30} z_{32}  z_{50}  z_{52} \sqrt{z_{31} z_{40} z_{42}z_{51}}}
\biggr[b z_{42} z_{51} \left\{ p q (p+q) z_{31} z_{32}^2 z_{43}-z_{30}^2 \left(q z_{52}^2+p \left(-2 z_{20}+z_{30}+z_{50}\right)
   z_{53}\right)\right\} a^3\\
&   +a^2 \left\{b^2 \left(q z_{21} z_{31} z_{32}^2 z_{42} z_{43} p^3+z_{42} \left(\left(z_{20}^2-z_{30} z_{50}+z_{10} \left(-2
   z_{20}+z_{30}+z_{50}\right)\right) z_{53} z_{30}^2+2 q^2 z_{21} z_{31} z_{32}^2 z_{43}\right) p^2 \right.\right.\\
& \quad \left.\left.   +p q \left(z_{10}^2 z_{40} z_{42} z_{32}^2+q^2 z_{21} z_{31} z_{42}   z_{43} z_{32}^2
   +z_{30} \left(z_{43} z_{20}^4+\left(4 z_{30}^2-z_{40} z_{30}-3 z_{40} z_{50}\right) z_{20}^3 
  -z_{30}^3z_{20}^2-\left(3 z_{40}+4 z_{50}\right)z_{20}^2 z_{30}^2\right.\right.\right.\right.\\
&\quad \quad  \left.\left.\left.\left.+6   z_{20}^2 z_{40} z_{50} z_{30}+z_{20}^2 z_{40} z_{50} \left(z_{40}+z_{50}\right)
+z_{30} \left(z_{40} z_{30}^2+z_{50} \left(z_{40}+2 z_{50}\right) z_{30}-2 z_{40} z_{50}
   \left(z_{40}+z_{50}\right)\right) z_{20}-z_{30}^2 z_{40} z_{50} z_{54}\right)\right.\right.\right.\\
&\quad\quad \left.\left.\left.   +z_{10} \left(\left(\left(2 z_{40}-z_{50}\right) z_{30}-3 z_{30}^2+3   z_{40} z_{50}\right) z_{20}^3
-z_{43} z_{20}^4-\left(z_{30}^3-6 z_{50} z_{30}^2+z_{40} \left(z_{40}+5 z_{50}\right) z_{30}+z_{40} z_{50} \left(z_{40}+z_{50}\right)\right)   z_{20}^2\right.\right.\right.\right.\\
&\quad\quad \left.\left.\left.\left.+z_{30} \left(z_{30}^3+\left(2 z_{40}-z_{50}\right) z_{30}^2+\left(2 z_{40}^2-3 z_{50} z_{40}-2 z_{50}^2\right) z_{30}+2 z_{40} z_{50}
   \left(z_{40}+z_{50}\right)\right) z_{20}+z_{30}^2 z_{40} \left(z_{54} z_{30}+z_{50} z_{54}-z_{30}^2\right)\right)
   \right) \right.\right.\\
&\quad \left. \left.
   +q^2 z_{20} z_{31} \left(z_{30}^2-2 z_{40}   z_{30}+z_{20} z_{40}\right) z_{52}^2+z_{30}^2 z_{52}^2 \left(z_{20} z_{53}-z_{40} z_{51}\right)\right)  -z_{30}^2 z_{42} z_{51} z_{52}^2+p q z_{31} z_{32}^2 z_{42}   z_{43} z_{51}\right\}  \\
&   +b \left\{z_{31} z_{40} \left(p q (p+q) z_{21} z_{42} z_{32}^2+\left(q z_{20}^2-2 q z_{30} z_{20}-p z_{30}^2\right) z_{52}^2\right) b^2+q z_{31}
   \left(z_{40} z_{20}^2-2 z_{30} z_{43} z_{20}-z_{30}^2 z_{40}\right) z_{52}^2\right.\\
&\quad \left.   +2 p^2 q z_{21} z_{31} z_{32}^2 z_{42} z_{43}+p z_{42} \left(\left(\left(z_{50}-2   z_{30}\right) z_{20}^2+2 z_{30} z_{50} z_{20}+z_{30} \left(z_{30}-2 z_{50}\right) z_{50}\right.\right.\right.\\
&\quad \quad \left.\left.\left.+z_{10} \left(z_{20}^2+2 \left(z_{30}-2 z_{50}\right) z_{20}-z_{30}^2+2
   z_{50}^2\right)\right) z_{30}^2+2 q^2 z_{21} z_{31} z_{32}^2 z_{43}\right)\right\} a\\
&   -z_{30}^2 z_{31} z_{42} z_{52}^2+p q z_{21} z_{31} z_{32}^2 z_{42} z_{43}+b^2 z_{31}
   z_{40} \left(p q z_{21} z_{32}^2 z_{42}-z_{30}^2 z_{52}^2\right)\biggr],
\end{split}
\end{align}
\begin{align}
\begin{split}
&\hat{\kappa}^{-1}= \frac{ v_0^2 \sqrt{2z_{41}} }{(a b (p+q)+1)^2 z_{30} z_{32}  z_{50}  z_{52} \sqrt{z_{31} z_{40} z_{42}z_{51}}}
\biggr[b z_{51} \left\{z_{40} \left(p q z_{30} z_{20}^3-p q z_{30} \left(4 z_{30}+z_{40}\right) z_{20}^2+p q z_{30}^2 \left(z_{30}+2
   z_{40}+4 z_{50}\right) z_{20}\right.\right.\\
&\quad \left.\left.   -p q z_{30}^3 z_{40}+p q \left(z_{31} (p+q)^2+z_{10}\right) z_{32}^2 z_{42}-z_{30}^2 \left(2 p q z_{50}^2+\left(p^2+q^2-1\right)
   z_{52}^2\right)\right) b^2\right.\\
&\quad \left.  +z_{30}^2 z_{42} z_{52}^2-p^2 z_{30}^2 z_{32}^2 z_{42}   -p q z_{32}^2 z_{42} \left(z_{30} z_{40}-z_{10} z_{43}\right)\right\} a^3\\
&   +a^2\left\{b^2
   \left(p^2 q z_{31} z_{42} \left(-z_{10} z_{30}-z_{20} z_{40}+\left(z_{30}+z_{40}\right) z_{50}\right) z_{32}^2\right.\right.\\
&\quad \left.\left.   -q \left(z_{30} \left(z_{40} z_{20}^2+z_{30}
   \left(z_{50}-2 z_{40}\right) z_{20}+z_{30} z_{40} z_{50}\right)-z_{10} \left(z_{40} z_{20}^2+z_{30} \left(z_{30}-2 z_{40}\right) z_{20}+z_{30}^2 z_{40}\right)\right)
   z_{52}^2\right.\right.\\
&\quad \left.\left.   +p \left(q^2 z_{31} z_{32}^2 z_{42} \left(-z_{10} z_{30}-z_{20} z_{40}+\left(z_{30}+z_{40}\right) z_{50}\right)-z_{30}^2 \left(\left(2
   \left(z_{40}-z_{50}\right) z_{50}-z_{30} \left(z_{40}-2 z_{50}\right)\right) z_{20}^2\right.\right.\right.\right.\\
&\quad \left.\left.\left.\left.   +z_{50} \left(z_{50} \left(z_{50}-2 z_{40}\right)-z_{30}^2\right) z_{20}+z_{40}
   z_{50} \left(z_{30}^2-z_{50} z_{30}+z_{50}^2\right)\right.\right.\right.\right.\\
&\quad \left.\left.\left.\left.   -z_{10} \left(\left(2 z_{30}+z_{40}-2 z_{50}\right) z_{20}^2+\left(-z_{30}^2-2 z_{40} z_{30}+z_{50}^2\right)
   z_{20}+z_{30}^2 z_{40}\right)\right)\right)\right)-p z_{30}^2 z_{32}^2 z_{42} z_{51}\right\} \\
&   +b \left(z_{31} z_{40} \left(-q^2 z_{52}^2 z_{32}^2-p q
   \left(z_{50}^2-z_{10} z_{42}+z_{20} \left(z_{40}-2 z_{50}\right)\right) z_{32}^2+z_{30}^2 z_{52}^2\right) b^2+z_{30}^2 \left(z_{30} z_{40}-z_{10} z_{42}-z_{20}
   z_{50}\right) z_{52}^2\right.\\
&\quad \left.   +p q z_{31} z_{32}^2 z_{42} \left(-z_{20} z_{40}+z_{10} z_{43}+z_{30} z_{50}\right)\right) a-b^2 q z_{31} z_{32}^2 z_{40} z_{52}^2\biggr].
\end{split}
\end{align}

\subsection{${\cal G}_i$ and ${\cal H}_i$}\label{app:cfs}
\begin{align}
\begin{split}
& {\cal G}_0 =  -12 \nu ^6 \left(\nu ^2-1\right), \quad
 {\cal G}_1 =  \nu ^4 \left(2 \nu ^6+\nu ^4 (15-16 n)+4 \nu ^2 (6 n-1)-21\right), \\
& {\cal G}_2 =  \nu ^2 \left(-4 \nu
   ^8+39 \nu ^6-98 \nu ^4+43 \nu ^2-4 \left(8 \nu ^4-13 \nu ^2+3\right) n^2+\left(44 \nu ^6-6 \nu ^4-44 \nu ^2+6\right) n+12\right),\\
   & {\cal G}_3 =  -\nu ^{10}-72 \nu ^8+40
   \nu ^6+117 \nu ^4-41 \nu ^2-16 \left(\nu ^2-1\right) \nu ^2 n^3+\left(4 \nu ^8+10 \nu ^6+47 \nu ^4-24 \nu ^2+3\right) n^2\\
   &\qquad-2 \left(\nu ^6-12 \nu ^4+65 \nu ^2-14\right)   \nu ^2 n-3, \\
&   {\cal G}_4 =  6 \nu ^{10}-13 \nu ^8+138 \nu ^6-35 \nu ^4-18 \nu ^2+8 \left(\nu ^6-5 \nu ^4+5 \nu ^2-1\right) n^3+\left(-12 \nu ^8+82 \nu ^6+21 \nu ^4-20 \nu   ^2+1\right) n^2\\
&\qquad+\left(\nu ^{10}-67 \nu ^8-14 \nu ^6-116 \nu ^4+41 \nu ^2-5\right) n+10, \\
& {\cal G}_5 =  -3 \nu ^{10}+76 \nu ^8-56 \nu ^6-14 \nu ^4+63 \nu ^2+4 \left(\nu   ^2-1\right)^2 n^4-8 \left(\nu ^6-3 \nu ^4+\nu ^2+1\right) n^3\\
&\qquad+\left(13 \nu ^8-50 \nu ^6+107 \nu ^4-40 \nu ^2+26\right) n^2-2 \left(\nu ^{10}-25 \nu ^8+82 \nu ^6-33 \nu
   ^4+27 \nu ^2+8\right) n-2,\\
&    {\cal G}_6 =  -31 \nu ^8+34 \nu ^4+16 \nu ^2+4 \left(\nu ^2-1\right)^2 n^4+16 \left(\nu ^2-1\right) n^3+\left(-3 \nu ^8+22 \nu ^6-63 \nu   ^4+48 \nu ^2+12\right) n^2\\
&\qquad+\left(\nu ^{10}-13 \nu ^8+72 \nu ^6-40 \nu ^4-53 \nu ^2+1\right) n-3,
   \end{split}
     \end{align}
     \begin{align}
        \begin{split}
&{\cal H}_0 =  -12 \nu ^8 \left(\nu ^2-1\right),\quad {\cal H}_1 =  \nu ^6 \left(3 \nu ^6-8 \nu ^4 (2 n+1)+3 \nu ^2 (8 n+9)-30\right),\\
&{\cal H}_2 =  2 \nu ^4 \left(23   \nu ^6-41 \nu ^4-5 \nu ^2-2 \left(8 \nu ^4-13 \nu ^2+3\right) n^2+\left(19 \nu ^4-15\right) \nu ^2 n+15\right),\\
&{\cal H}_3 =  -\nu ^2 \left[9 \nu ^{10}+11 \nu ^8+108   \nu ^6-190 \nu ^4+15 \nu ^2+16 \left(\nu ^2-1\right) \nu ^2 n^3+\left(-8 \nu ^8+29 \nu ^6-112 \nu ^4+49 \nu ^2-6\right) n^2\right.\\
&\qquad\left.-2 \left(36 \nu ^8-29 \nu ^6-53 \nu ^4+3 \nu
   ^2+3\right) n+15\right] ,\\
&   {\cal H}_4 =  -91 \nu ^{10}+187 \nu ^8+150 \nu ^6-130 \nu ^4+9 \nu ^2+16 \left(\nu ^6-5 \nu ^4+5 \nu ^2-1\right) \nu ^2 n^3-4 \left(6 \nu
   ^{10}-40 \nu ^8-3 \nu ^6+8 \nu ^4+\nu ^2\right) n^2\\
   &\qquad+\left(4 \nu ^{12}-97 \nu ^{10}+37 \nu ^8-290 \nu ^6+90 \nu ^4+19 \nu ^2-3\right) n+3,\\
   &{\cal H}_5 =  6 \nu ^{12}+53
   \nu ^{10}+164 \nu ^8-84 \nu ^6-19 \nu ^4+33 \nu ^2+8 \left(\nu ^2-1\right)^2 \nu ^2 n^4-8 \left(\nu ^2-1\right)^2 \left(\nu ^2+3\right) \nu ^2 n^3  \\
   &\qquad +\left(85 \nu ^8+92
   \nu ^6-60 \nu ^4+11\right) n^2-\left(57 \nu ^{10}+149 \nu ^8+186 \nu ^6-158 \nu ^4+37 \nu ^2+9\right) n-1,\\
   &{\cal H}_6 =  73 \nu ^{10}-59 \nu ^8-4 \nu ^6+72 \nu ^4-\nu
   ^2+16 \left(\nu ^2-1\right)^2 \nu ^2 n^4+4 \left(\nu ^{10}-11 \nu ^8+22 \nu ^6-10 \nu ^4+\nu ^2-3\right) n^3\\
   &\qquad +2 \left(24 \nu ^{10}-47 \nu ^8+84 \nu ^6-64 \nu ^4+34 \nu
   ^2+5\right) n^2+\left(-6 \nu ^{12}+51 \nu ^{10}-227 \nu ^8+112 \nu ^6-24 \nu ^4-59 \nu ^2+1\right) n-1,\\
   &{\cal H}_7 =  \nu ^2 \left(\nu ^{10}-28 \nu ^8+2 \nu ^6+32 \nu
   ^4+13 \nu ^2-4\right)+4 \left(\nu ^2-1\right)^2 \left(\nu ^4+1\right) n^4-4 \left(\nu ^{10}-5 \nu ^8+12 \nu ^6-16 \nu ^4+7 \nu ^2+1\right) n^3\\
&\qquad   +\nu ^2 \left(\nu ^{10}-10
   \nu ^8+73 \nu ^6-112 \nu ^4+32 \nu ^2+32\right) n^2-\left(35 \nu ^{10}-79 \nu ^8+20 \nu ^6+52 \nu ^4+5 \nu ^2-1\right) n.
   \end{split}
     \end{align}

\end{document}